\documentclass{emulateapj}
\usepackage[usenames,dvipsnames,svgnames,table]{xcolor}
\usepackage{hyperref}
\definecolor{darkblue}{rgb}{0.0,0.0,0.3}
\hypersetup{colorlinks,breaklinks,
            linkcolor=darkblue,urlcolor=darkblue,
            anchorcolor=darkblue,citecolor=darkblue}
\usepackage{amsmath,amssymb}
\usepackage{graphicx,verbatim}
\usepackage{amsmath}
\usepackage{amssymb}
\usepackage{bm}
\usepackage{natbib}

\begin{document}
\def\etal{et al.\ \rm}
\def\ba{\begin{eqnarray}}
\def\ea{\end{eqnarray}}
\def\etal{et al.\ \rm}
\title{Ab-initio pulsar magnetosphere: the role of general relativity.}
\author{Alexander A. Philippov\altaffilmark{1}$\dagger$, Beno\^it Cerutti\altaffilmark{1}$\ddagger$, Alexander Tchekhovskoy\altaffilmark{2}$^\ast$
\& Anatoly Spitkovsky\altaffilmark{1}}
\affil{$^1$ Department of Astrophysical Sciences, 
Princeton University, Ivy Lane, Princeton, NJ 08544}
\affil{$^2$ Departments of Physics and Astronomy, University of California, Berkeley, CA 94720}
\altaffiltext{$\dagger$}{sashaph@princeton.edu}
\altaffiltext{$\ddagger$}{Lyman Spitzer, Jr. Fellow}
\altaffiltext{$\ast$}{Einstein Fellow}


\begin{abstract}

It has recently been demonstrated that self-consistent particle-in-cell simulations of low-obliquity pulsar magnetospheres in flat spacetime show weak particle acceleration and no pair production near the poles. We investigate the validity of this conclusion in a more realistic spacetime geometry via general-relativistic particle-in-cell simulations of the aligned pulsar magnetospheres with pair formation. We find that the addition of frame-dragging effect makes local current density along the magnetic field larger than the Goldreich-Julian value, which leads to unscreened parallel electric fields and the ignition of a pair cascade. When pair production is active, we observe field oscillations in the open field bundle which could be related to pulsar radio emission. We conclude that general relativistic effects are essential for the existence of pulsar mechanism in low obliquity rotators. 



\end{abstract}

\keywords{acceleration of particles -- magnetic fields -- plasmas -- pulsars: general}




\section{Introduction}

Physical phenomena in pulsar magnetospheres can now be studied with first-principles plasma simulations using particle-in-cell (PIC) codes. Last several years saw significant progress in modeling various regimes of magnetospheric plasma supply. \citet{PS14} showed that when pairs are injected throughout the magnetosphere of the aligned rotator so that abundant plasma is present everywhere, the system approaches the force--free state (also see \citealt{Belyaev15}).  \citeauthor{Benoit15} (\citeyear{Benoit15}; hereafter, CPPS15) reached the same conclusion when injecting quasineutral plasma at the stellar surface; they also studied particle acceleration in the current sheet. \citeauthor{Chen14} (\citeyear{Benoit15}; hereafter, CB14) incorporated pair production into the aligned pulsar simulation, and \citeauthor{P15} (\citeyear{P15}; hereafter, PSC15) extended this study to oblique rotators.

New challenges appeared as well. It was shown that there is no pair formation in the bulk of the polar cap in the aligned rotator (CB14) and in the oblique magnetospheres with inclination $\alpha\lesssim 40^\circ$ (PSC15). In these solutions the current density along the magnetic field, $j_{\parallel}$, is below the Goldreich--Julian (GJ) value, $j_{GJ}=-\vec{\Omega}_* \cdot \vec{B}/ 2\pi$ (here, $\vec{\Omega}_*$ is the stellar angular velocity vector, and $\vec{B}$ is the local magnetic field vector), and the parallel electric field is screened by a mildly relativistic charge-separated flow of particles lifted from the neutron star (NS) surface \citep{Shibata97, Bel08, Tim13}. This slow outflow does not radiate gamma-photons needed to produce pairs.  The lack of polar cap activity in pulsars with low inclination angles seems implausible from an observational standpoint, which implies that some important physics is missing from the models. In particular, PSC15 suggested that effects of general relativity (GR), which are significant in the primary particle acceleration region \citep{B90, MT92, ShibataGR}, may increase the local value of $j_{\parallel}/j_{GJ}$ and lead to the ignition of discharge.

In this Letter we study the effects of general relativity on the aligned pulsar magnetosphere. To estimate the strength of these effects, we first construct GR force-free simulation of the aligned rotator in $\S$\ref{sec:ffree}. In $\S$\ref{sec:GRPIC} we describe our implementation of GR in the PIC method. In $\S$\ref{sec:Results} we present the GR PIC simulation of the aligned pulsar magnetosphere with pair production and show that the polar pair cascade is restored in GR. The implications of these findings are described in \S\ref{sec:conclusions}.

\section{GR force-free simulation of the aligned rotator}
\label{sec:ffree}


The spacetime around a rotating NS can be described using the Kerr metric, which is parametrized by stellar mass $M$ and spin $a$.  This assumption ignores corrections due to non-uniform stellar mass distribution. 
To estimate the importance of GR effects,  we performed the simulation  of the aligned magnetosphere in Kerr metric using the force-free version of \texttt{HARM} code \citep{HARM,WHAM, AAA15}. 
As in \citet{McKinney06}, we set the inflow velocity into the current sheet to zero, in order to minimize the numerical dissipation. In our simulation the stellar compactness parameter is fixed to $r_s/R_*=0.5$, where $r_s=2GM/c^2$ is the Schwarzschild radius.  The spin parameter in units of stellar radius is set to $a/R_* = 2/5 (\Omega_* R_*/c)$ for a spherically symmetric uniform NS interior. We run the simulation for a range of ratios of stellar to light cylinder radii: $0.02< R_*/R_{LC}< 0.2$, where $R_{LC} \equiv c/\Omega_*$. The initial magnetic field configuration corresponds to dipolar field in the Schwarzschild metric:
\begin{equation}
A_{\phi}=\frac{3\mu\sin^2\theta}{r_s}\left(\frac{1}{2}+\frac{r}{r_s}+\left(\frac{r}{r_s}\right)^2\ln{\left(1-\frac{r_s}{r}\right)}\right),\label{GRdip}
\end{equation}
where $A_{\phi}$ is the $\phi$ component of the vector potential, and $\mu$ is the stellar dipole moment as observed from infinity.

\begin{figure}
\hspace*{-0.1cm}
\includegraphics[width=0.5\textwidth]{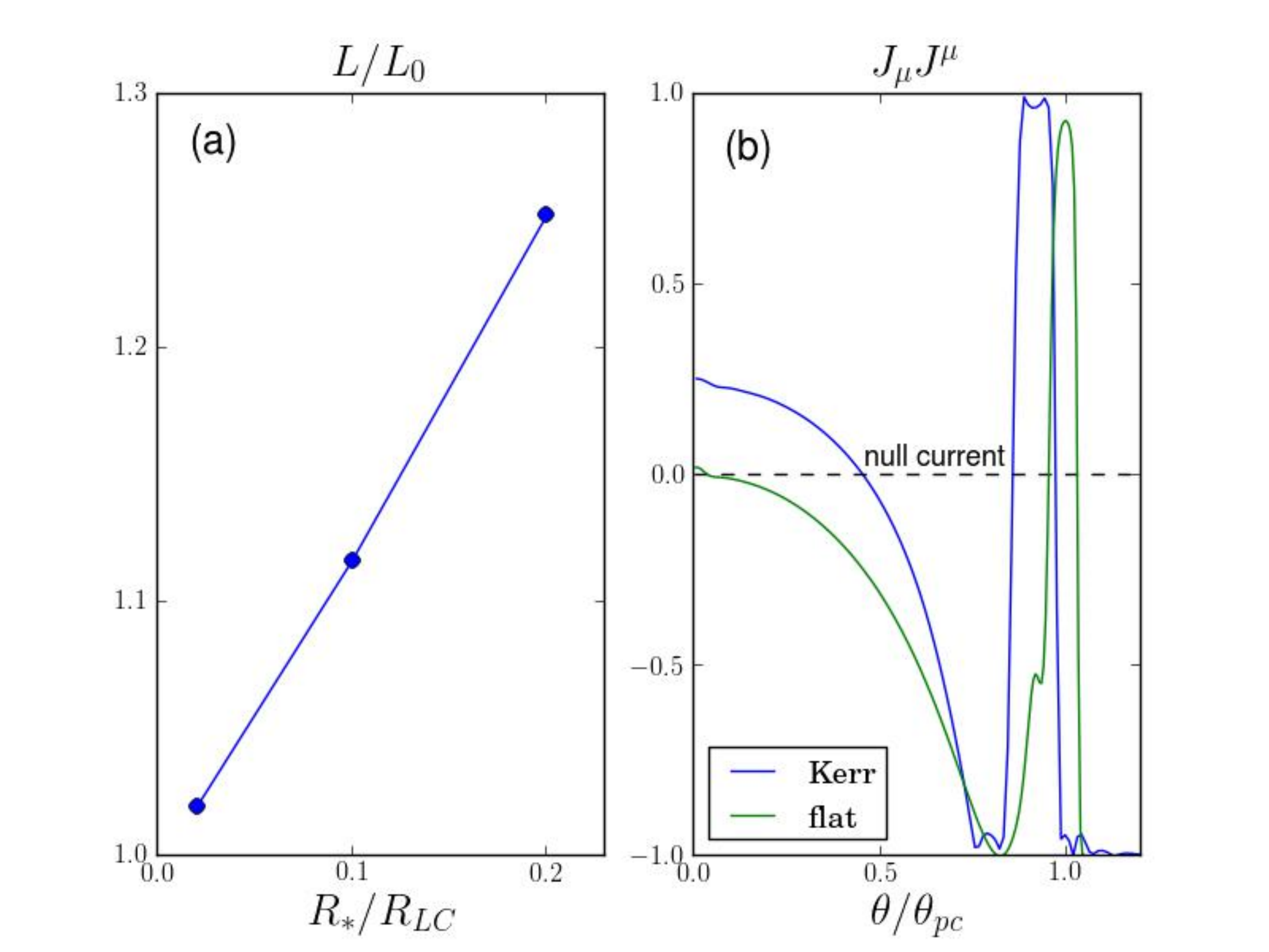}
\caption{Force-free simulation of GR aligned pulsar magnetosphere. (Panel a): Spindown luminosity $L$ increases with increasing stellar to light cylinder radius ratio. (Panel b): Square of the 4--vector of current distribution $J^2\equiv{}J^\mu J_\mu$ on the polar cap; $\theta_{pc}$ is the polar cap boundary. GR effects cause the current to become space-like at the rotational axis.}
\label{fig1:ff}
\end{figure}

The steady state magnetosphere is similar to the flat-space solution, with open and closed field lines, Y-point and the current sheet. The spindown power is enhanced by up to $20\%$ compared to the flat-space value\footnote{We find that the GR enhancement is due to the larger value of the open magnetic flux.}, $L_{0}=\mu^2\Omega_{*}^4/c^3$, consistent with \citet{Ruiz14}. This correction disappears in the limit of slow rotation, $R_*/R_{LC} \to 0$ (see Figure \ref{fig1:ff}a), applicable for ordinary radio pulsars. 

We calculate the current distribution $J^{\mu}=F^{\mu\nu}_{;\nu}$ on the polar cap and find that the current is space-like ($J^{\mu}J_{\mu} \geq 0$) on a large fraction of the polar cap (see blue line in Figure \ref{fig1:ff}b), regardless of the $R_*/R_{LC}$ ratio. Since in the force-free model charge density closely follows GJ density, this conclusion is equivalent to $J_{\hat{r}}/\rho c \approx {j_{\parallel}}/\rho_{GJ}c \geq 1$. This can be understood as follows. Since frame-dragging effectively reduces the angular velocity of the star, GJ charge density at the pole is also reduced compared to the flat-space value \citep{B90, MT92}:
\begin{equation}
\rho \approx \rho_{GJ}\approx-\frac{(\Omega_{*}-\omega_{LT})B_*}{2\pi c \alpha},
\end{equation}
where $\alpha$ is the lapse and $\omega_{LT}$ is the Lense-Thirring angular frequency at the polar cap. However, current density in the polar region is set by the twist of magnetic field lines at the light cylinder, which is located far enough from the star to be unaffected by GR effects. Thus, 
\begin{equation}
\frac{J_{\hat r}}{\rho_{GJ} c}\approx \left(\frac{J_{\hat r}}{\rho_{GJ} c}\right)_{\textrm{flat}}\frac{1}{1-\omega_{LT}/\Omega_*}.
\label{expect}
\end{equation}
Since the current is null (the parenthesized value in eq.~\ref{expect} is unity) on the rotational axis in flat-spacetime (green line in Figure \ref{fig1:ff}b), we expect a finite polar region with $J_{\hat r} /\rho_{GJ}c > 1$ in GR. To screen the electric fields on these field lines, pair production is required \citep{Bel08}. This conclusion is dramatically different from the flat-space solution, which has time-like currents over most of the polar cap, except for the rotational axis and thin return current layer at the edge of the open flux tube (Figure \ref{fig1:ff}b, also \citealp{Parfrey12}). The flow on field lines with time-like current may be sustained by particles with one sign of charge, which is the reason for the lack of pair formation in pulsars of low obliquities in flat space.



\section{GR PIC method Implementation}
\label{sec:GRPIC}
We now describe a GR extension \texttt{EZeltron} of the 2.5D (axisymmetric) PIC code \texttt{Zeltron}, which was earlier applied to pulsar magnetospheres (CPPS15). Since $r_s$ is located well below the NS surface, the calculation can be performed in Boyer-Lindquist coordinates. In the slow rotation limit, $a/r_s \ll 1$, the metric is \citep{MT92}
\begin{align}
{\rm d}s^2= \alpha^2(c{\rm d}t)^2-\left(\frac{{\rm d}r^2}{\alpha^2}+r^2 ({\rm d}\theta^2  +  \sin^2\theta  {\rm d}\phi^2)\right)  +  \nonumber\\2g_{0i}c{\rm d}t{\rm d}x^i = \alpha^2(c {\rm d} t)^2 - \sigma_{ij}x^i x^j + 2g_{0i}c{\rm d}t{\rm d}x^i,
\end{align}
where ($t, r, \theta, \phi$) are the time, radial and angular spherical coordinates, $\alpha=\sqrt{1-r_s/r}$,    $\vec{g}_0=\vec{\beta} = \vec{\omega}_{LT}\times \vec{r}/c$, and $\sigma_{ij}$ is the spatial part of the Schwarzschild metric.  For a spherically symmetric NS the Lense-Thirring angular frequency is \begin{equation}\omega_{LT}=\frac{2}{5}\Omega_* \frac{r_s}{R_*}\left(\frac{R_*}{r}\right)^3.\end{equation}At the stellar surface for typical compactness $r_s/R_{*}\approx 0.5$, its value is comparable to the stellar spin $\omega_{LT}/\Omega_*\approx 0.2$.

In 3+1 formalism, the particle equations of motion are \citep{Thorne, Fisch}
\begin{align}
\label{particle motion} \frac{{\rm d}{\vec p}}{{\rm d}t} &= \alpha q (\vec{E} + \frac{{\vec v}}{c} \times {\vec B}) + \alpha m\gamma {\vec g} + \alpha \overset{\text{\tiny${\bm\leftrightarrow}$}}{\rm{H}} {\vec p},\\
\frac{{\rm d} \vec x}{{\rm d} t} &= \alpha \vec{v} - \vec{\beta}
\label{particle advance}
\end{align}
where $q$ is particle charge, $\vec{p}=m\vec{v}\gamma$ is particle momentum, $\vec{g}$ is the gravitational acceleration, and $\overset{\text{\tiny${\bm\leftrightarrow}$}}{\rm{H}}$ is the gravitomagnetic tensor. For very strong pulsar magnetic fields the last two terms in Eq. \ref{particle motion} may be justifiably neglected. Maxwell equations in 3+1 splitting are \citep{Thorne}
\begin{align}
\label{poisson eq} \nabla \cdot \vec{E} &= 4\pi \rho,\\
\nabla \cdot \vec{B} &= 0, \\
\nabla \times \left(\alpha\vec{E} +\frac{\vec{\beta}}{c}\times \vec{B}\right)
	&= - \frac{1}{c}\frac {\partial{\vec B}} {\partial t}, \label{faraday eq}\\
\label{induct eq} \nabla \times \left(\alpha\vec{B} - \frac{\vec{\beta}}{c}\times \vec{E}\right)
	&= \frac{1}{c}\frac {\partial{\vec E}} {\partial t} +\alpha \vec{j} - \rho\vec{\beta},
\end{align}
In \texttt{EZeltron} we solve the particle equation of motion (\ref{particle motion}) neglecting gravity forces using the Boris algorithm and advance particle positions according to (\ref{particle advance}). Maxwell's equations (\ref{faraday eq})-(\ref{induct eq}) in curved space are solved on axisymmetric $(r,\theta)$ Yee lattice. Physical vector components of particle velocities $v_{\hat{i}}=\sqrt{\sigma_{ii}(v^i)^2}$ and electromagnetic fields $E_{\hat{i}}, B_{\hat{i}}$ are used in the code loop. To solve Maxwell's equations, differential operators are integrated over a cell (CPPS15, \citealt{Belyaev15}), whose length and area are modified in the Schwarzschild metric. Particle charges and currents are deposited on the grid using the volume weighting technique (CPPS15), accounting for GR corrections to cell volumes, $V_{cell} = \int 2\pi\sqrt{-\sigma} {\rm d} r{\rm d}\theta$. Our current deposition scheme is not strictly charge-conservative; thus, we correct the electric field by solving equation (\ref{poisson eq}) using the iterative Gauss--Seidel method.

Our simulations start in vacuum with a GR dipolar magnetic field (derived from eq. \ref{GRdip}) frozen into the surface of a conducting NS. At $t=0$ the star is spun up to its angular velocity $\Omega_*$ by imposing electric field\begin{align}
E_{\hat{\theta}} &= -(\Omega_*-\omega_{LT})B_{\hat{r}} R_* \sin \theta/\alpha c,\\
E_{\hat{\phi}} &= 0,
\end{align}at the stellar surface. To prevent possible reflections, we put an absorbing layer for both fields and particles at the outer boundary of the simulation domain (CPPS15).

\section{Results}
\label{sec:Results}
\subsection{Test: monopole solution}

\begin{figure}
\hspace*{-0.8cm}
\includegraphics[width=0.5\textwidth]{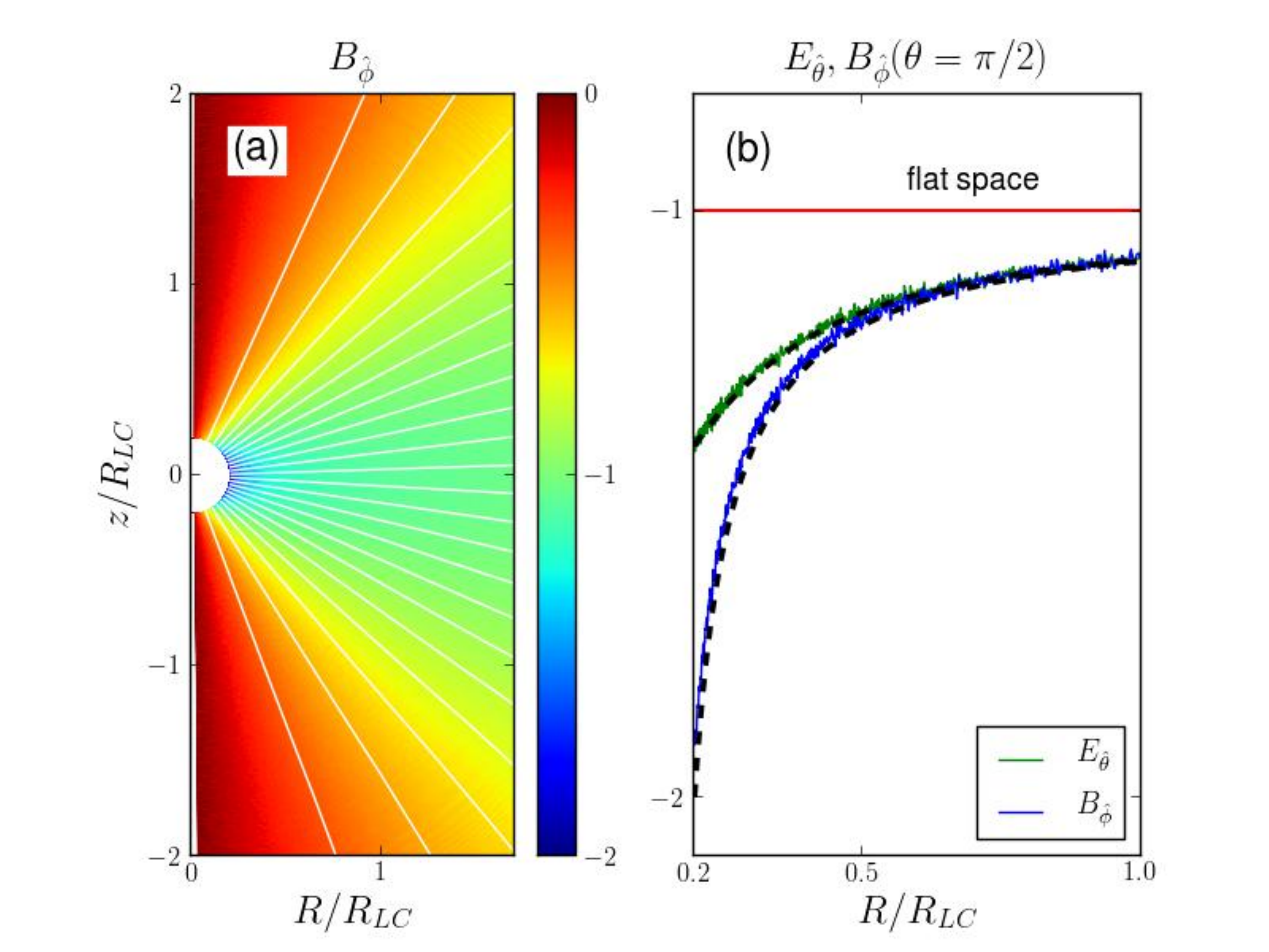}

\caption{Monopole test with GR PIC. Electromagnetic fields are normalized to $(r/R_{LC})B_{\hat{r}}$. (Panel a): Poloidal field lines are shown as white, color represents toroidal field $B_{\hat{\phi}}$. (Panel b): Slice of $B_{\hat{\phi}}$ and $E_{\hat{\theta}}$ at the equator. Black dashed lines show analytical estimates of $B_{\hat{\phi}}$ (\ref{monopole:Schw}) and $E_{\hat{\theta}}$ (\ref{monopole:frame-drag}). Red solid line represents flat spacetime solution $E_{\hat{\theta}}=B_{\hat{\phi}}$.}
\label{fig2:mono}
\end{figure}
As a test case we describe the monopolar magnetosphere calculated with PIC, for which an analytical solution is known in the Schwarzschild geometry \citep{Lyutikov}:\begin{equation}E_{\hat{\theta}} = B_{\hat{\phi}} = - \frac{\Omega_* r\sin\theta}{c\alpha} B_{\hat{r}}.\label{monopole:Schw}\end{equation}This solution represents the case of the null current $J_{\hat{r}}=\rho{}c=\Omega_*B_{\hat{r}}\cos\theta/2\pi\alpha$. 

We construct a plasma-filled solution by injecting charges of both signs at the surface of the star at a constant rate moving along the field lines with a radial velocity of 0.5c (CPPS15). When we set $\omega_{LT} = 0$, we reproduce the analytical solution (\ref{monopole:Schw}). With frame-dragging, we find that $B_{\hat{\phi}}$ and the current density $J_{\hat{r}}$ remain mostly the same (see Fig. \ref{fig2:mono}). However, the corotating electric field $E_{\hat{\theta}}$ is reduced, being approximately\begin{equation}E_{\hat{\theta}}\approx -\frac{(\Omega_*-\omega_{LT})r\sin\theta}{c\alpha}B_{\hat{r}}.\label{monopole:frame-drag}\end{equation}Thus, the ratio $J_{\hat r}/\rho c \approx B_{\hat{\phi}}/E_{\hat{\theta}}$ increases in agreement with (\ref{expect}). The current becomes space--like, $J_{\hat r}/\rho c>1$, in the region where frame--dragging is efficient, i.e., $R_*\leq r \lesssim 2R_*$. 

\subsection{Aligned magnetosphere with pair production}

\begin{figure*}
\includegraphics[width=0.98\textwidth]{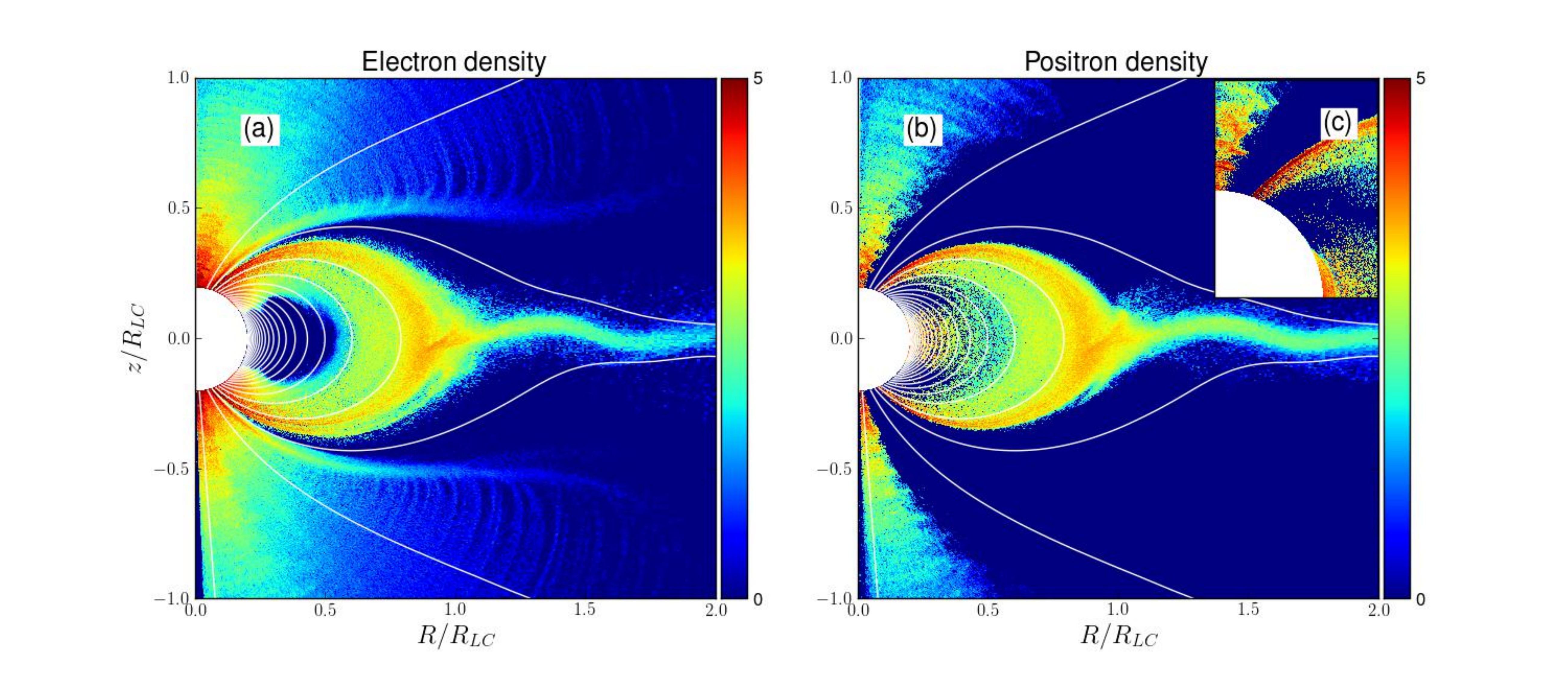}
\caption{The inclusion of GR effects triggers pair production in an aligned pulsar magnetosphere, which is shown after four rotational periods for $r_s/R_*=0.7$. Color represents plasma densities on the logarithmic scale, normalized to GJ density at the pole. White lines show magnetic field lines. (Panel a): Electron density. (Panel b): Positron density. (Panel c): Zoom-in onto the positron density close to the star.}
\label{fig3:dipole}
\end{figure*}

\begin{figure}
\includegraphics[width=0.5\textwidth]{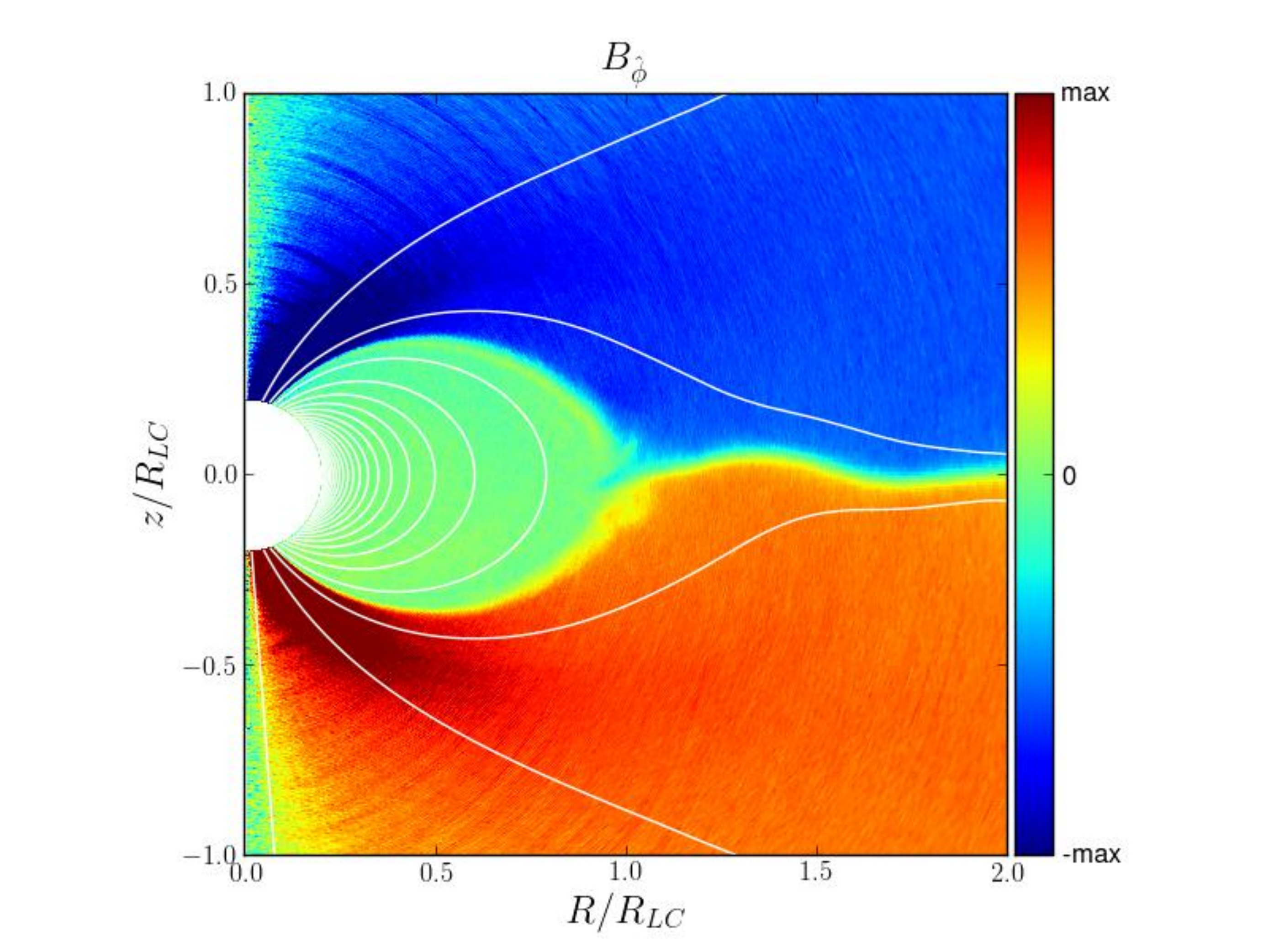}
\caption{Aligned pulsar magnetosphere with pair production, shown after four rotational periods, $r_s/R_*=0.7$. Color represents toroidal component of the magnetic field $B_{\hat{\phi}}$. White lines show magnetic field lines. Field oscillations are due to time-dependence of polar discharge and could be related to pulsar radio emission (see text for details).}
\label{fig4:dipole}
\end{figure}

In this section we present the structure of the aligned rotator with GR effects, calculated with PIC. Our treatment of plasma supply closely follows PSC15, which we summarize below for completeness. In order to mimic particle extraction from the NS surface, we inject neutral plasma at rest with the star half a cell below the stellar surface, with the rate controlled by the amount of surface charge available: $0.1(E_{\hat{r}}(r_*) - E_{\hat{r}}^{cor}(r_*))/4\pi$, where $E_{\hat{r}}^{cor}$ is the radial component of corotation electric field. This injection results in a non-neutral plasma emerging from the surface. Particles are removed from the simulation domain if their radial position is one cell below the stellar surface. If particles accelerate above the star, they can cross the threshold for pair formation, $\gamma_{min} = 20$, which we choose to be independent of the distance from the star. The threshold is small compared to the Lorentz factor that a particle would gain from the full pole-to-equator potential drop, $\gamma_0 = (\Omega_* R_*/c) (e B_* R_*/m_e c^2) \approx 2000$. Secondary particles are created at locations of primary energetic particles and are injected with Lorentz factor of 4 in the direction of motion of primary particles. We put a density limiter for produced pairs, so that the local multiplicity of the flow does not exceed 8. The stellar radius in our simulation is $R_*=0.2 R_{LC}$, and our computational grid extends to $4 R_{LC}$ in the radial direction. We also apply the radiation reaction force on particles, which accounts for both synchrotron and curvature radiation \citep{Tamburini10}. Our grid points are uniformly spaced in $\log_{10} r$ and $\theta$ (CPPS15), the grid size is $4096 \times 1024$ cells, so $R_{LC}$ is resolved with 2200 cells in the radial direction. The skin depth at the stellar surface is resolved with 15 cells for plasma with multiplicity 8 and $\gamma=1$.

Apart from not having polar cap pair production, our steady-state magnetosphere is very similar to the aligned solution of CB14 and PSC15: pair formation is active in the return current layer and the equatorial current sheet. With the GR effects taken into account, we find that pair formation at the pole is activated when $r_s/R_* \gtrsim 0.5$. This is because the current density near the polar axis becomes larger than the local GJ value. In this case the current can not be sustained by the primary electron beam alone, so both charge species must be present. Figure \ref{fig3:dipole}a and \ref{fig3:dipole}b show the distribution of electrons and positrons in the magnetosphere. In Figure \ref{fig3:dipole}c we plot the positron distribution near the stellar surface, which shows non-stationary behavior of  pair production on the polar cap. Pair plasma produced in the active phase of the discharge gets polarized and screens $E_{\parallel}$, so the pair formation stops. As the pair cloud leaves the gap region, the accelerating voltage is restored, and the cycle repeats \citep{Tim13}. The period of the discharge intermittency is slightly larger than $h/c$, where $h$ is the gap height. In our simulation the gap height is determined by the distance that a particle travels before it is accelerated up to the the pair formation threshold, which scales roughly as $h \sim R_{pc} (\gamma_{min}/\gamma_{pc})$. Here, $R_{pc}$ and $\gamma_{pc}$ are the polar cap width and potential drop. As in CB14 and PSC15, increasing the threshold for pair formation up to the full polar cap potential value leads to the suppression of  return current, and the magnetosphere relaxes to  electrostatically-confined disk-dome solution. 

The non-stationary pair production produces an imprint on the structure of electromagnetic field, which is shown in the distribution of $B_{\hat{\phi}}$ in Figure  \ref{fig4:dipole} (there are similar oscillations of $E_{\hat{\theta}}$). The  current and charge density fluctuations responsible for these oscillations are produced in the process of screening of $E_{\parallel}$ during the discharge, which happens on plasma skin scale, while the period between successive pulses is determined by the discharge intermittency $\sim h/c$ \citep{Bel08}. We expect that the waves that we observe in the open field line region are directly connected with the pulsar radio emission. The energy in these fluctuations, $W \approx 10^{-2}L_0$, is more than enough to power the observed radio activity. Cascade simulations with more detailed physics of pair formation and realistic multiplicities infer somewhat smaller wave amplitudes, $W \approx 10^{-4} L_0$ \citep{Tim13}, though only electrostatic waves were possible in their 1D study. Future work should study the fraction of the wave energy escaping from the magnetosphere in the form of transverse electromagnetic oscillations. We note that these waves are absent in the flat space simulation, which has no pair production at the polar cap.

Accounting for radiation reaction force does not influence the global magnetospheric structure (in particular, the current distribution at the polar cap), but makes the current layers thinner in response to the loss of pressure support \citep{U11}. The pair production is more localized: without radiation losses primary particle Larmor gyrations are larger in size, and many secondary particles are injected with non-zero angle with respect to the current layer, creating an atmosphere. In the presence of radiation losses particle orbits around the sheet are more focused, and secondary pairs are mostly injected along the direction of the current flow. We find that the drift-kink instability of the current sheet is less violent in the presence of radiative cooling, though it is still prominent by the end of our simulations (see Figure~\ref{fig3:dipole} for plasma density).

\subsection{Stellar Compactness and Pair Production}
\label{sec:stell-comp-pair}

As seen in Figure~\ref{fig3:dipole}, whereas the polar cascade injects abundant plasma along the polar magnetic field lines, it leaves a vacuum-like gap near the equatorial current sheet. In this case the volume-distributed return current, which flows in this region in force-free solutions, is suppressed. Since the polar cap current closes via the return current near the current sheet, this reduces the maximum flat-space polar cap current to $(J_{\hat{r}}/\rho_{GJ}c)_{\rm{}flat}\approx0.8$ down from $1$ in the force-free solution. In order for the polar cascade to operate, $J_{\hat{r}}/\rho_{GJ}{}c$ needs to exceed unity, requiring $\omega_{LT}/\Omega_*\gtrsim 0.2$, or stellar compactness exceeding a critical value, $r_s/R_*\gtrsim 0.5$. If an additional source of plasma in the gaps near the current sheet is present, this can enable the volumetric return current, increasing $(J_{\hat{r}}/\rho_{GJ}c)_{\rm{}flat}$ to unity. Then, we would expect the requirement of critical compactness to be relaxed.

In order to testп this, we try the following experiment in flat space. We run the simulation until the magnetosphere reaches a steady state, at 1.5 rotational periods, and inject neutral plasma in every cell that has electron density smaller than the local GJ value, at a rate of $2n_{GJ}$ per rotaional period. We find that the gap is filled with quasi-neutral plasma, the current in the polar cap region increases and reaches the force--free value $J_{\hat{r}} \approx \rho_{GJ} c$ on the axis. This is accompanied by the increase of the spindown energy losses, from $0.8 L_0$ to $L_0$, consistent with the plasma-filled solutions of \citet{PS14} and CPPS15. Thus, if the gap region is filled with plasma and frame--dragging is taken into account, the current density on field lines close to the polar axis is expected to become larger than the local GJ value at much smaller values of $r_s/R_*$ (see Eq. \ref{expect}), leading to a discharge. Pairs of low multiplicity in the gap region may be produced in collisions of thermal X-ray photons from the NS surface upscattered by energetic particles in the magnetosphere, curvature photons in the return current layer or synchrotron photons from the current sheet. On the other hand, the gap has non-zero $E_{\parallel}$, and it remains to be seen if active discharge in this region can be ignited by particles produced in photon collisions.

\section{Conclusions}
\label{sec:conclusions}
We presented the magnetospheric structure of an aligned rotator with pair production, accounting for the effects of general relativity. We found that for solutions with compactness $r_s/R_*>0.5$, pair formation is active in the part of the polar cap, proceeding in an essentially non--stationary way. We observed field oscillations in the polar cap region that are the direct consequence of a time--dependent discharge, rather than the maser-type amplification process in the steady flow due to plasma instabilities, as is usually assumed. We speculate that these waves should be the primary source for the observed coherent pulsar radio emission. We find that the compactness value needed to activate pair production corresponds to a star with $R_*=10\,\rm{km}$ and $M_*=1.7M_{\odot}$. When the vacuum gap above the current sheet is filled with neutral plasma, the ratio $J_{\hat{r}}/\rho_{GJ}c$ in the polar cap region increases, and this should help creating polar discharge at lower compactness values. We will explore this possibility in future work. According to the estimate (\ref{expect}), the critical compactness should be lower for oblique rotators since the ratio $J_{\hat{r}}/\rho_{GJ} c$ increases with inclination in flat space. 

We conclude that the effects of general relativity are essential for the operation of pulsar polar cap discharge. In particular, in pulsars with low inclination angle the high multiplicity of pair plasma and the observed radio emission are a direct consequence of frame-dragging. For pulsars with intermediate inclinations,  taking into account GR effects will likely activate pair production on a larger fraction of the polar cap than was found in PSC15. This is particularly important for stars with small $R_{*}/R_{LC}$ ratio where the null charge surface does not cross the polar cap. 

We thank Jonathan Arons, Vasily Beskin, Sam Gralla and Andrey Timokhin for fruitful discussions. This research was supported by NASA Earth and Space Science Fellowship Program (grant NNX15AT50H to AP), NASA grants NNX14AQ67G, NNX15AM30G, Einstein Postdoctoral Fellowship (grant PF3-140115 to AT) awarded by the Chandra X-ray Center, which is operated by the Smithsonian Astrophysical Observatory for NASA under contract NAS8-03060, Simons Foundation (through Simons Investigator grant to AS), and was facilitated by Max Planck/Princeton Center for Plasma Physics. The simulations presented in this article used computational resources supported by the PICSciE-OIT High Performance Computing Center and Visualization Laboratory.

\end{document}